\RequirePackage{fix-cm}

\documentclass{article}
\usepackage{comment}

\usepackage[nearskip=-3pt,captionskip=4pt,
            listofformat=subsimple,
            labelformat=simple]{subfig}

\usepackage{graphicx}

\usepackage{pict2e}

\usepackage{amsthm}
\usepackage{amssymb,amsfonts,amsmath}
\usepackage{fourier}

\newtheorem{theorem}{Theorem}

\newcommand{\mvec}{{\bf m}}
\newcommand{\nvec}{{\bf n}}
\newcommand{\jhalf}{\textstyle{\frac{1}{2}}}

\newcommand{\jpoint}[1]{\ell_{{#1}}}

\begin{document}
\bibliographystyle{plain}
\title{The  Fundamental Theorem of  Phyllotaxis revisited}
\author{Jonathan Swinton \\ Deodands Ltd\\ \texttt{jonathan@swintons.net}}
\maketitle

\begin{abstract}
\textit{(July 2024): This paper has now been superseded and, whilst still mathematically correct, should be considered obsolete.
}\\[2em]
Jean's `Fundamental Theorem of Phyllotaxis'
(\emph{Phyllotaxis: a systematic study in Plant Morphogenesis}, CUP 1994)
describes the 
relationship between the count numbers of observed spirals in 
cylindrical lattices and the horizontal angle between vertically successive
spots in the lattice. It is indeed fundamental to observational studies of
phyllotactic counts, and especially to the evaluation of hypotheses about the 
origin of Fibonacci structure within lattices. Unfortunately the textbook version
of the theorem is incomplete in that it is incorrect for an important special case. This paper provides a complete statement and proof of the Theorem.
\end{abstract}

%
%

\section*{Note}
This paper was first uploaded to arXiv in 2012, and the counterexample here to Jean's theorem as he stated it remains valid. Since that time Infang Publishing has published my  book-length treatment of \textit{Mathematical Phyllotaxis}, ISBN 0993178960. This book  puts the (minor) significance of the problem in more context and supersedes this paper, which should be considered obsolete -- Jonathan Swinton, July 2024.

\section{Introduction}

Mathematical phyllotaxis is the study of the patterns that appear in two-dimensional cylindrical lattices, given particular motivation by the striking
appearance of high Fibonacci numbers in a range of biological settings such as the spirals on a sunflower~\cite{jean:1994}.  Although  static analyses of 
 lattices  cannot in themselves  explain the appearance of these numbers~\cite{turing:1992}, they are essential both  in relating what can actually be biologically observed to 
 hypothesises about the underlying order, and in forming a basis for dynamical models of lattice formation~\cite{mitchison:1977,douady:1992,smith:2006,atela:2011} that can,
it is claimed, explain 
Fibonacci numbers and related structure in biological form. 
More specifically, a phyllotactic theory of lattices creates a model for which lines in the lattice are most likely to be remarked on by a human observer. 
In the case of the sunflower or the fir cone, these lines may be those which join adjacent points in the lattice, which may be defined in different ways  as contact parastichies ~\cite{richards:1951,jean:1994} or principal parastichies~\cite{turing:1992}. 
A slightly more general idea is to identify those  pairs of lines that wind in opposite directions as opposed parastichies, or alternatively those lines which can
be thought of as characterising the lattice, which were defined as generating parastichies by Turing~\cite{turing:1992} or equvalently as visible parastichies by Jean~\cite{jean:1994}.

Jean presents the most complete description in the literature of  mathematical phyllotaxis in his textbook~\cite{jean:1994}, and deserves considerable credit both for innovation and
 a substantial work of integration, bringing together a range of biological datasets and historical mathematical approaches. One major contribution  is what he calls the \emph{Fundamental Theorem of Phyllotaxis} which he attributes in a special case to Adler~\cite{adler:1977}. The basic idea of the theorem is a very useful conceptual one.
Cylindrical lattices can be, for these purposes, completely characterised by the angle of rotation between successive points called the divergence.
If a lattice is seen to posses a specific pair of generating and opposed parastichies, characterised in a natural way by a pair of integers, then there is a constraint on the allowable values
of  the divergence. The theorem shows that for any pair of integers not both equal to 1 there  are exactly two  intervals of nonzero width on which 
the divergence will create the required generating and opposed parastichies. 
Unfortunately, however, in the form stated by Jean the Theorem needs modification in a range of special but  important cases. The goal of this note is to restate the Theorem completely. First we give the necessary background about two-dimensional cylindrical lattices, and give a characterisation of which divergence values correspond to a given generating pair.
After restating the Fundamental Theorem in Jean's formulation and demonstrating a counter example, we then reprove a suitably corrected Theorem. 

Unbeknown to Jean and Adler, Turing had also considered very similar problems, but this work was unpublished at his death in 1954~\cite{swinton:2003}. 
It remained accessible but obscure in the Turing Archive in King's College Cambridge, until being published in his collected works in 1992~\cite{turing:1992}, well after the relevant papers of Jean and Adler.  As a secondary aim, this paper points out the ways in which Turing anticipated the later, more widely known work.

\section{Background}

This section contains a number of statements without proof that are fairly obvious on examining a diagram. They can be made rigorous by eg the use of congruences~\cite{turing:1992}.

We consider  a cylinder of circumference 1 and extending infinitely in the vertical direction,
 with an origin and coordinates $(x,z)$, $0\leq x \leq 1$. For any $0\leq d\leq1$, we can construct a lattice $d$
by rotating by an angle $2\pi d$ around the cylinder from the origin and rising by $z=1$, and repeating. This creates the set of points  $(x,z)=(d_m,m)$ where $m$
is any integer and $d_m=md-[md]$ and $[x]$ is the nearest integer to $x$, so that $-\jhalf \leq d_m \leq \jhalf$.
Since taking $[\jhalf]=0$ and $[\jhalf]=1$ map to the same point on the cylinder we will allow the function $[x]$ to take the multiple values $0$, $1$  at the point $x=\jhalf$. 
We call $(d_m,m)$ the point $\jpoint{m}$, and from now on assume $m\neq 0$. The vertical component is called the \emph{rise}.

By construction we have excluded lattices with more than one point at each rise. More generally if there are $J$ such points spaced equally around the cylinder we would describe the
lattice as having \emph{Jugacy} $J$, but we restrict to $J=1$ here.

\begin{figure}
\includegraphics{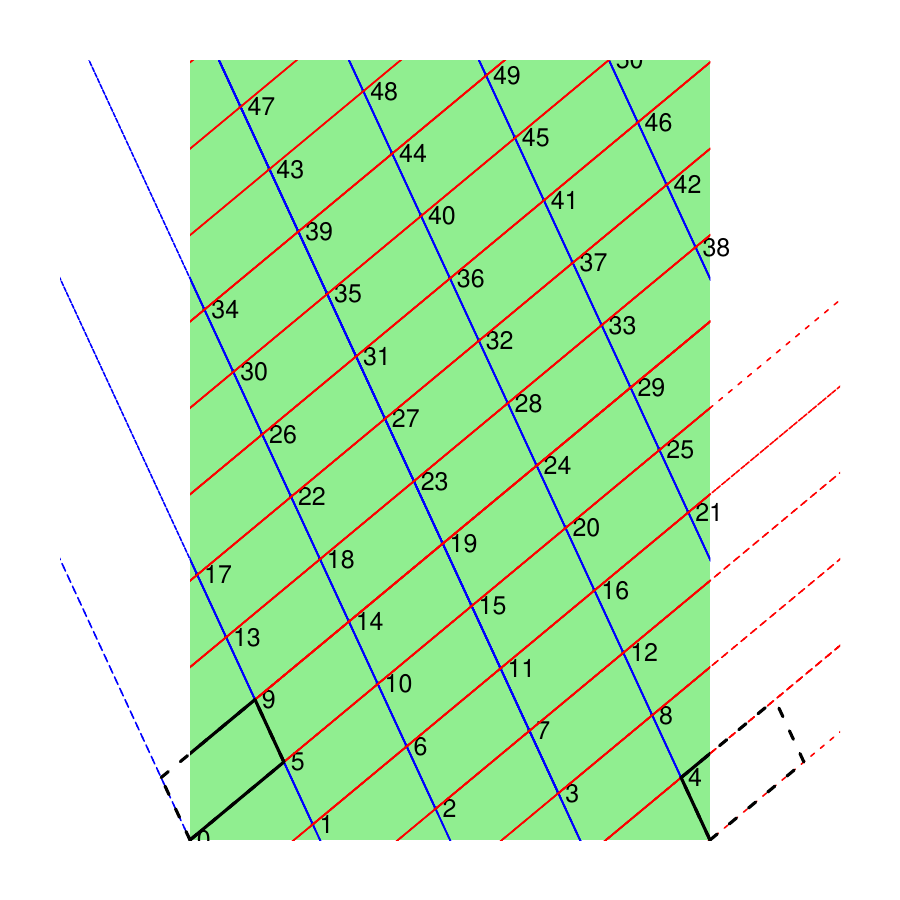}
\caption{(4,5) is a generating and opposed pair for
the cylindrical $d=17/72$ lattice. The parallelogram defined by the pair tiles the lattice and every lattice point 
is at a vertex of one of the parallelograms; the edges of the parallelograms form the parastichy lines.
The blue lines highlight the family of 4-parastichies and the red lines the family of 5-parastichies.
  }
\label{fig:plines45}
\end{figure}

A parastichy of order $m$ is the infinite line through the origin and $\jpoint{m}$ with slope $d_m/m$. 
There are two possible lines on the cylinder through $0$ and $\jpoint{m}$ corresponding to winding in opposite directions and this choice of slope is equivalent to choosing the line that traverses the smallest $x$ distance between $0$ and $\jpoint{m}$, leaving an ambiguity when $[md]=\jhalf$.
The portion of this line between $0$ and $m$ defines the vector $\mvec$, again  with an an ambiguity when $[md]= \jhalf$.  An $m$-parastichy is a member of the family of $m$ distinct lines containing the origin-parastichy of order $m$ and the parallel lines to it through the points $1,\ldots,m-1$.  If a point $\jpoint{p}$ is on an $m$-parastichy then so is $\jpoint{p+m}$. See Figure~\ref{fig:plines45}. 

A pair of (parastichy) numbers $(m,n)$ define a pair of points  $\jpoint{m}$,  $\jpoint{n}$ and vectors $\mvec,\nvec$.
 $(m,n)$ is \emph{opposed} if $d_m/m$ and $d_n/n$ have opposite sign. 
In the case when $[md]=\jhalf$ we define the parastichy pair $(m,m)$ as the pair combining each of the choices of direction around the cylinder.
There remains an ambiguity when $[md]=\jhalf$ or $[nd]=\jhalf$ and $m\neq n$ which could be resolved by a specific choice of direction although it is not of significance subsequently.

There is a natural relationship between a cylindrical lattice and a corresponding periodic lattice in
the plane, and the $m$-parastichies also define an infinite family of $m$-parastichies in the plane lattice. 
A pair is \emph{generating} for $d$ if it generates the lattice 
in the plane in the sense that every point can be expressed as a vector sum $v\mvec+u\nvec$ for integer $u,v$.%
\footnote{Note that the definition is such because  any non collinear vectors would generate the lattice \emph{in the cylinder}. }
It is necessary, but not sufficient, for a generating pair $(m,n)$ to be coprime for if they have a common factor $k$ all rises, including $1$ must be a multiple of $k$.
This is effectively the definition given by Turing~\cite{turing:1992}, and identical  to the  \emph{visible} pair defined by 
Jean~\cite{jean:1988,jean:1994} in a number of different equivalent ways. Since Jean~\cite{jean:1994} gives no proof of that equivalence
we give it in the Theorem below which also establishes the identity with the Turing definition, and in the process
modify some of Jean's definitions for extra precision. I have chosed to stick with Turing's word \emph{generating} over Jean's \emph{visible} for these 
identical concepts as I think the latter word carries confusing connotations in the identification of parastichy counts.

We make use of the determinant $\Delta_{mn}$ of a pair $(m,n)$, defined as
\begin{eqnarray*}
\Delta_{mn}(d) &=& \left[nd\right]m -  \left[md\right]n\\
&=&  ( n d - d_n)m -  ( m d - d_m)n \\
&=& n d_m - m d_n,
\end{eqnarray*}
 except for the special case $\Delta_{mm}(\jhalf)=m$ (where we have picked $[\jhalf]=1$ in the $[nd]$ and $[\jhalf]=0$ in the $[md]$).

\begin{theorem}
(Compare Theorem 4.2 of Jean~\cite{jean:1994}.)
The following are equivalent
\end{theorem}

\begin{enumerate}
\item The pair $(m,n)$ is generating in the lattice $d$. \label{lem:gen}
\item The pair $(m,n)$ has a point of the lattice $d$ at every intersection of the lines of the pair. \label{lem:int}
\item The points $0$,  $\jpoint{m}$, and $\jpoint{n}$ form a nondegenerate triangle which contains no other point of the lattice $d$  internally. \label{lem:tri}
\item  $(m,n)$ satisfy $|\Delta_{mn}(d)|=1$\label{lem:delta}. 
\end{enumerate}

\begin{proof}

If a pair of vectors are collinear in the plane, they cannot be
generating. If they are not collinear, the parallelogram formed by any pair
can be used to tile the cylinder. This tiling will contain lattice points exactly
 at the vertices of each parallelogram iff the pair is generating, because if it has an internal point it must be a noninteger sum of the pair.
This shows~\ref{lem:gen}$\Leftrightarrow$\ref{lem:tri}. Moreover the tiling produces the parastichy families of order $m$ and $n$, so these must always intersect at a lattice point iff
the pair is generating. This shows~\ref{lem:gen}$\Leftrightarrow$\ref{lem:int}. 

A pair is generating iff it can express the 
unit vector $(0\leq d< 1,1)$ as a sum of $\mvec$ and $\nvec$ in the plane.
In plane coordinates, we have
\begin{eqnarray*}
\left[nd\right]\mvec-\left[md\right]\nvec &=& \left( \left[nd\right]  d_m - \left[md\right]  d_n , \left[nd\right] m - \left[md\right] n \right)
\\
&=& \left( \left[nd\right]\left(md-\left[md\right]\right) - 
	\left[md\right]\left(nd-\left[nd\right]\right), \Delta_{mn} \right)\\
&=& \Delta_{mn} (d, 1)
\end{eqnarray*}
If $\Delta_{mn}=1$ we are done, and if $\Delta_{mn}=-1$ we take the combination $\left[md\right]\nvec-\left[nd\right]\mvec$ 
of opposite sign, so if $|\Delta_{mn}|=1$ then $(m,n)$ is generating. This shows~\ref{lem:delta}$\implies$\ref{lem:gen}.

To prove~\ref{lem:gen}$\implies$\ref{lem:delta}, the central idea (of Jean and Adler and Turing) is to continue
the $m$ and $n$ parastichies away from the origin until they cross again, so first we have to dispose of the case when the two parastichies
are parallel.
If $\mvec$ and $\nvec$ are parallel, then they are not generating, and moreover $d_m/m=d_n/n$ so $\Delta_{mn}=0$ and conversely.
If they are not parallel then in the plane the parastichy of order $m$ through $(0,0)$ and $(d_m,m)$
and and the parastichy of order $n$ through $(1,0)$ and $(d_n,n)$ must meet at the point 
\[
\left(
	\frac{md_n}{\Delta_{mn}}
	=
	\frac{nd_m}{\Delta_{mn}}-1
	,
	\frac{mn}{\Delta_{mn}}
\right).
\]
If $(m,n)$ is generating this must be a point of the lattice and so have rise equal to both $km$ and $k'n$ for integer $k,k'$, so
$n=k\Delta_{mn}$ and $m=k'\Delta_{mn}$. But since $m$ and $n$ are coprime,  $|\Delta_{mn}|=1$.

\end{proof}
There are close connections with the theory of Farey sequences, as mentioned in Jean~\cite{jean:1994} and in more detail in Jean~\cite{jean:1988}, which can be exploited to give different versions of this proof, but we do not pursue that here.

The existence of two choices for $\Delta$ is a reflection of the symmetry arising from the choice of direction
around the cylinder which corresponds to $(m,n,d,\Delta)\rightarrow(m,n,1-d,-\Delta)$ and $(m,n,d,\Delta)\rightarrow(n,m,d,-\Delta)$, so it is possible  to force
at least one of $m\leq n$ or  $|d|<\jhalf$ or $\Delta=+1$ if we wish. Indeed Jean chooses to focus in the case $d<\jhalf$, but here
we allow either choice but  recognise that the resulting intervals for $d$ are related by this symmetry.

\begin{figure}
\centering
\subfloat[(5,9) is a generating but not opposed pair.]{
\includegraphics[scale=0.5]{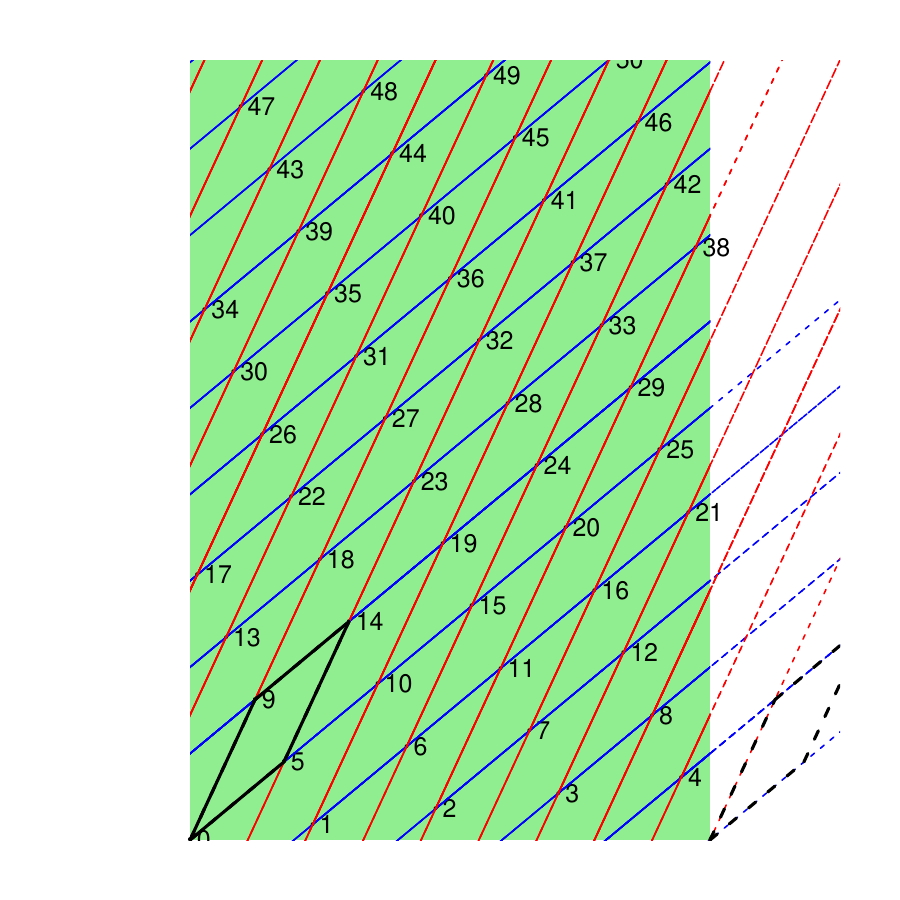}
\label{fig:plines59}
}
\subfloat[(9,19) is a nonopposed nongenerating pair which is not collinear.]{
\includegraphics[scale=0.5]{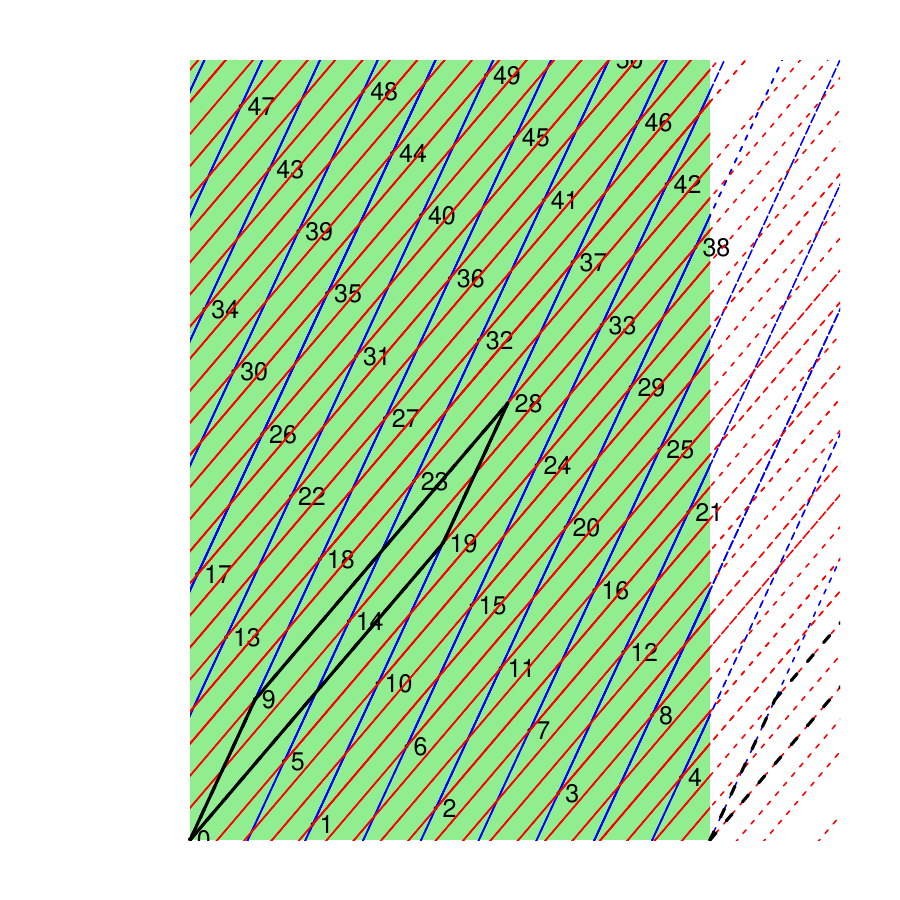}
\label{fig:plines919}
}
\\
\subfloat[(5,7) is  an opposed but not generating pair. ]{
\includegraphics[scale=0.5]{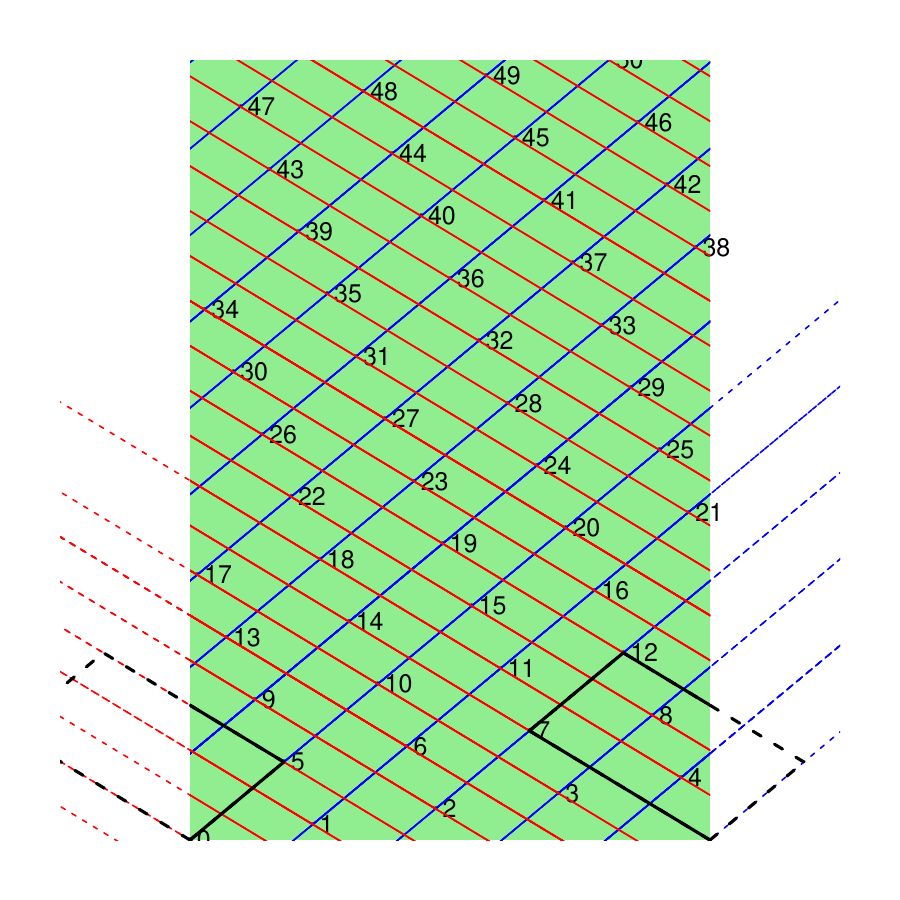}
\label{fig:plines57}
}
\subfloat[(1,2) is a nonopposed nongenerating  pair which is collinear. ]{
\label{fig:plines12}
\includegraphics[scale=0.5]{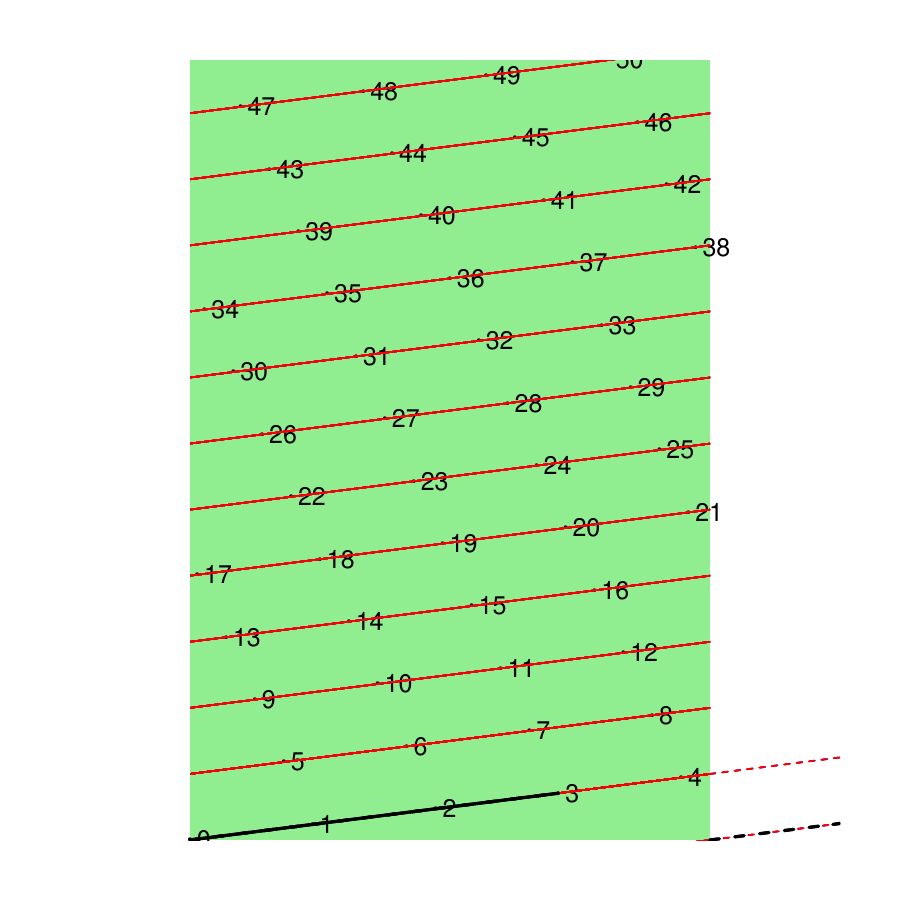}
}
\caption{Different types of parastichy pair in the lattice with divergence $d=17/72$.}
\label{fig:plinesex}
\end{figure}

Figure~\ref{fig:plines45} shows a generating opposed pair, and Figure~\ref{fig:plinesex} show a variety of pairs which are not.

The Fundamental Theorem gives conditions for $d$ if $(m,n)$ are generating and opposed. We will prove it
by first finding conditions for $(m,n)$ to be generating.

\section{Finding $d$ given $(m,n)$ generating}
\label{sec:g}
The previous Theorem gives only an implicit form for $d$. Here we find the explicit intervals for $d$
on which $|\Delta_{mn(d)}|=1$. Given $m,n$ coprime and $\Delta=\pm1$ we want to find those  $d$  such that $\Delta_{mn}(d)=\Delta$.

If $m=n$  but $\mvec\neq\nvec$ we must have $d=\jhalf$ and we are done, with  $m=n=1$, $\Delta=1$. 

Otherwise, assume for now that $m<n$.
Now take  $u$, $v$ by solving $mv-nu=\Delta$, specified uniquely for $m>1$ by $0\leq u < m$, $0\leq v <n$,
 or for $m=1$ by $(u,v)=(0,1)$ when  $\Delta=1$ or $(u,v)=(1,n-1)$  when $\Delta=-1$. 

To force $[md]=u$ and $[nd]=v$ we need 
\begin{eqnarray*} 
L_m = \frac{u-\jhalf}{m} \leq &d& \leq \frac{u+\jhalf}{m} = R_m
\\
L_n = \frac{v-\jhalf}{n} \leq &d& \leq \frac{v+\jhalf}{n} = R_n
\end{eqnarray*} 
respectively and $d$ is in the intersection of the intervals $(L_n,R_n)$, $(L_m,R_m$). Note that eg $(L_n,R_n)$ is centred at $v/n$ and has width $1/n$.
Then
\begin{eqnarray*}
 mn (L_n-L_m) &= &         \Delta +\jhalf (n-m)  
\\
mn (R_n- R_m) &=&          \Delta -\jhalf (n-m)  
\\
mn (R_n-L_m) &=& \Delta +\jhalf(m+n)
\\
mn (R_m-L_n) &=& -\Delta+\jhalf(m+n)
\end{eqnarray*}

So $L_n>L_m \Leftrightarrow n>m-2\Delta$,
which is always true unless $\Delta=-1$ and $n=m+1$. 
Similarly $R_m>R_n $ iff $n>2\Delta+m$ which is true unless $\Delta=1$ and $n=m+1$.
So apart from those two cases we have $L_m<L_n<R_n<R_m$ and the interval $(L_n,R_n)$ is the one we want.
\newcommand{\sign}{\mathrm{sign}}
To pay attention to the special cases we see 
that for $n=m+1$ 
\begin{eqnarray*}
\sign (L_n-L_m) &= & \sign(        \Delta +\jhalf  ) = \Delta
\\
\sign (R_n- R_m) &=&         \sign( \Delta -\jhalf ) = \Delta
\\
\sign (R_n-L_m) &=& \sign(\Delta +n+\jhalf) = +1
\\
\sign (R_m-L_n) &=& \sign(-\Delta+n+\jhalf) = +1
\end{eqnarray*}
so if $\Delta=1$ we have $L_m<L_n<R_m<R_n$ while if $\Delta=-1$ it is $L_n<L_m<R_n<R_m$.

We originally assumed $m<n$. If instead $m>n$, we can swap $m$ and $n$ which will change the sign of $\Delta$, so we can summarise
in 
\begin{theorem}
$(m,n)$ is a generating pair in the lattice $d$ iff $d$ is in the intervals specified in Table~\ref{tab:intervals}. 
\label{lab:thmg}
\end{theorem}

\begin{table}
\[
\begin{array}{c|cc}
& \Delta=1 &\Delta = -1
\\
\hline
n\leq  m-2 & (L_m, R_m)&(L_m,R_m)
\\
n =m-1 & (L_n,R_m)  & (L_m,R_n)
\\
n=m(=1)& \jhalf & 
\\
n=m+1 & (L_n,R_m)  & (L_m,R_n)
\\
n\geq m+2 & (L_n, R_n)&(L_n,R_n)
\end{array}
\]
\caption{Intervals on which $(m,n)$ is generating in $d$.}
\label{tab:intervals}
\end{table}
Armed with Theorem~\ref{lab:thmg} we can now add the additional condition that the parastichy pair be opposed in order to find the
Fundamental Theorem.

\section{The Fundamental Theorem of Phyllotaxis}

\subsection{The Jean formulation}
Jean's version~\cite{jean:1988,jean:1994} of the FTP states

\emph{
Let $(m,n)$ be a parastichy pair, where $m$ and $n$ are relatively prime, in a system with divergence angle $d$. The following properties are equivalent:
\begin{itemize}
\item[(1)] There exist unique integers $0\leq v<n$, amd $0\leq u<m$ such that $|mv-nu|=1$ and $d<\jhalf$ is in the closed interval whose end points are $u/m$ and $v/n$;
\item[(2)] The parastichy pair $(m,n)$ is visible and opposed
\end{itemize}
}
\subsection{Counterexample}
\begin{figure}
\begin{center}
\includegraphics{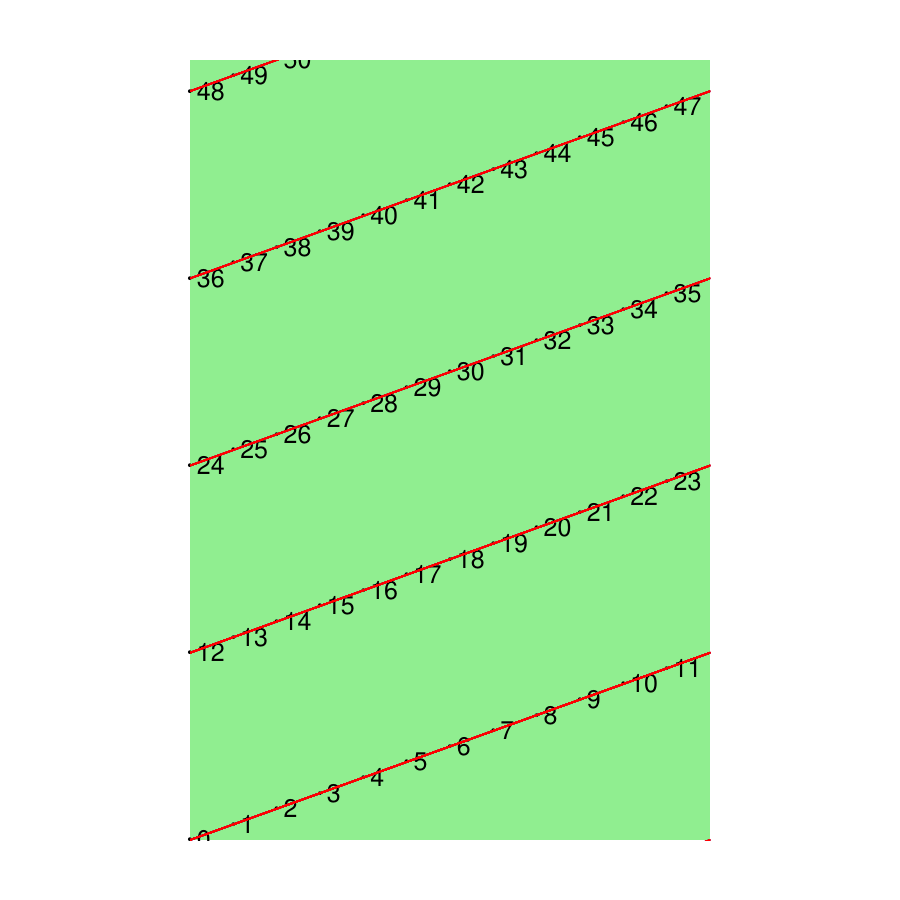}
\caption{(1,3) are not a generating or an opposed pair for the lattice $d=1/12$. }
\label{fig:plinescounterx}
\end{center}
\end{figure}

Consider the lattice with $d=1/12$. Then the parastichy pair $(m,n)=(1,3)$ is neither generating nor opposed, so (2) is false.
However the  unique integers satisfying $|mv-nu|=1$ and $0\leq u<1$ and $0\leq v<3$ are  $u=0$ and  $v=1$,  and $d=1/12<\jhalf$ is in the interval $[0/1,1/3]$, so (1) is true.
See Figure~\ref{fig:plinescounterx}.

In fact the counterexample holds for all pairs of  the form $(1,n)$, and in fairness to Jean it might be argued that $1$ and $n$ might not considered to be coprime, in which case the Theorem still holds, but this  interpretation is ruled out by the comment in Appendix 4.1  that non coprime pairs are those that do not produce `one genetic spiral'. 
A more powerful defense of the utility of the Jean version  of the Theorem is that it is intended for pattern recognition, typically on specimens with large, usually Fibonacci, parastichy numbers
in which it is only exceptionally the case that $m=1$. But even discounting the difficulty this error in the special case  can cause the reader in following the argument, it turns out that  all modern discussions of the appearance of Fibonacci structure~\cite{mitchison:1977,kunz:1995,atela:2002} invoke a successive sequence of bifurcations from more simple starting conditions,  specifically $(1,1)$ and $(1,2)$, so it is important to account properly for this case. 

\section{Opposed generating pairs}

We now need to reprove the FTP, which we do by considering on which portion of the $d$ interval where $(m,n)$ is generating
it is also opposing.  First we assume $\Delta>0$.

$d_m$ passes through $0$ at $L_m$, $u/m$, and $ R_m$ and nowhere else in 
the interval $(L_m,R_m)$, where $u$ is defined in the proof of Theorem~\ref{lab:thmg}. For $m>2$, by Theorem~\ref{lab:thmg} the generating interval is $(L_n,R_n)$ and 
we saw $L_m<L_n<u/m<v/n<R_n<R_m$, and so within the generating interval $d_m$ is negative only for $d<u/m$ and $d_n$ is negative only for $d<v/n$.
Thus the only region of the generating interval on which $(m,n)$ is opposed is $u/m<d<v/n$. Under the symmetry, we see the analogous case for $\Delta=-1$.

The point of this paper, though, is to define the  necessary interval when $m=1$. For $n=1$ we have already seen we must take $d=\jhalf$. 
Otherwise Theorem~\ref{lab:thmg} shows the generating interval (for $\Delta=+1$) is of the form $(1/4,1/2)$ for $n=2$ and  $1/n\pm 1/2n$ for $n>2$.
Since $d_n$ changes sign every $1/2n$, in either case,  the generating and opposed interval for $d$ is $(1/2n,1/n)$.
So we can summarise in 
\begin{theorem} (The Fundamental Theorem of Phyllotaxis). 
 The following are equivalent
\begin{enumerate}
\item $(m\leq n,n)$ is generating and opposed in the lattice $d$, with $\Delta_{mn}(d)=\Delta$
\item
\begin{enumerate}
\item $m=1$, $n=1$, $d=\jhalf$, and $\Delta=1$, or
\item $m=1$, $n>1$,   $\Delta=+1$, $d\in(1/2n,1/n)$, or
\item  $m=1$, $n>1$,  $\Delta=-1$,  $d\in 1 -(1/2n,1/n)$ , or
\item $1<m<n$, $d\in(u/m,v/n)$, $\Delta=\pm 1$,  where $u,v$ are the unique integers $0\leq v<n$, and $0\leq u<m$ such that $mv-nu=\Delta$.
\end{enumerate}
\end{enumerate}
\end{theorem}
Part of the significance of this theorem, as Adler~\cite{adler:1977} and Turing~\cite{turing:1992} showed, is that if $m=F_k$ and $n=F_{k+1}$ are successive members of the Fibonacci sequence, then the interval for $d$ is $(F_{k-2}/F_{k},F_{k-1}/F_{k})$ which rapidly converges to the point $d=\tau^{-2}$ where $\tau$ is the golden ratio.

\section{Discussion}

This correction to the Fundamental Theorem of Phyllotaxis does not reduce its centrality in the relationship between
observed parastichy counts and the underlying mathematical structure of cylindrical lattices. Nor does recasting it 
partly in the earlier work of Turing remove the justifiable priority claims of Jean and Adler in its development, since that earlier work was languishing unpublished and incomplete in the Turing archive
when they independently published theirs. Nevertheless, this paper has taken advantage of the correction needed to the special case when one of the parastichy numbers is 1 in order to 
put the Theorem in a more accurate historical context  and point out the common ideas of these authors.

\section{Acknowledgement}
I'm grateful to Paul Glendinning for comments on this draft.

\bibliography{js229}
\end{document}